\documentclass[preprintnumbers,amsmath,amssymb,amsfonts]{revtex4}

 \usepackage{amsmath,amssymb,latexsym,epsfig,graphics,epsf,bm}
\usepackage{graphics}
\usepackage{float}
\usepackage{placeins}
 \usepackage[toc,page]{appendix}

\begin{document}

   \title{Isomorph theory prediction for the dielectric loss variation along an isochrone}
\author{Wence Xiao\footnote{wence@ruc.dk}, Jon Tofteskov, Troels V. Christensen, Jeppe C. Dyre, and Kristine Niss}
\affiliation{DNRF Centre ``Glass and Time'', IMFUFA, Department of Sciences, 
Roskilde University, Postbox 260, DK-4000 Roskilde, Denmark}
\date{\today}
\begin{abstract}
This paper derives a prediction for the variation of the amplitude of 
the dielectric loss from isomorph theory, and presents an experimental 
test of the prediction performed by measuring the dielectric-relaxation 
behavior of the van der Waals liquid 5-phenyl-4-ether (5PPE).  The 
liquid is studied at isochronal states in the temperature range 
$266-333$ K and pressure range $0.1-300$ MPa, for relaxation times 
around $10^{-3}$ s and $10^{-4}$ s.  From the isomorph statement that 
there is structural and dynamic invariance of isomorph states in 
reduced units for Roskilde simple liquids we derive four equivalent 
isomorph-invariant terms, one of which is used in analyzing our 
data.  It is the frequency-dependent term $\chi_{e}(f) \rho^{\gamma-1}$, 
with electric susceptibility $\chi_{e}$, density $\rho$, and 
density-scaling factor $\gamma$. 

Due to the unique design of our experimental setup, we obtain dielectric 
loss data where the amplitude is reproducible $\pm 0.1 \%$.  We moreover 
find that the empty capacitance of the capacitor cell is stable within 
$\pm 0.3 \%$ in our measuring range and can be assumed to be constant.  Using 
this we predict for two isomorph states there is $-C_{2}''(f)= -C_{1}''(f) 
\left( \rho_1 / \rho_2 \right)^{\gamma-1}$ to scale the negative imaginary 
capacitance, where $C_{1}$ is the capacitance measurement at ambient 
pressure and $C_{2}$ is the predicted capacitance at elevated 
pressure.  We visually compare the predicted and measured plots and there 
is good match between the two plots among the 42 pairs of isochronal states 
from the measurement.

\end{abstract}

\maketitle

\section{Introduction}
The isomorph theory deals with a group of liquids called Roskilde
simple liquids \cite{whatisasimpleliquid}, which is developed through a 
series of papers \cite{pressureEnergy1,pressureEnergy2,pressureEnergy3,
pressureEnergy4,pressureEnergy5}.  It has been predicted from computer 
simulations that van der Waals liquids are generally considered 
simple \cite{Stronglycorrelatingliquids}.
 
For Roskilde simple liquid there are curves in the phase diagram along 
which structure and dynamics are invariant in reduced units 
\cite{pressureEnergy4}.  If state 1 and 2 are two states on the same 
isomorph curve, they have proportional Boltzmann factor for physical 
relevant microstates, and the proportionality is constant and state 
dependent.  The reduced units bring the two states with different densities 
into unit density.  Thus two isomorph states have one-to-one correspondence 
between their microscopic configurations \cite{pressureEnergy4}.  From 
this point we can derive some isomorph-invariant predictions for different 
physical properties.  

The isomorph theory can explain density scaling 
\cite{AlbaDensityScaling2002,RolandDensityScaling2005,Stronglycorrelatingliquids} 
and isochronal superposition \cite{superposition1,superposition3,Stronglycorrelatingliquids}, 
which have been experimental observations made from high-pressure dielectric
measurements before the development of the theory. It has moreover
been shown that isomorph theory can lead to non-powerlaw density
scaling and this behavior is likewise confirmed experimentally \cite{LasseDensityGamma} 
in line with earlier results suggesting non-powerlaw density
scaling \cite{KNfragility,Tarjus2004a,Targus2004b}.

The isomorph theory has proven very strong in predicting the behavior
of computer-simulated liquids.  A lot of the predictions from isomorph
theory which are tested by computer simulation are difficult to test
experimentally.  They refer to microscopic properties that are not
experimentally accessible or they require very large density changes
which are difficult (or impossible) to obtain in experiments
\cite{lassegamma}.

Among the work done for density scaling there is one previous paper
which tests a prediction of the isomorph theory without fitting
or scaling parameters \cite{dittepaper}.  It shows for a silicon oil
the density-scaling exponent
$\gamma_{scale}$ is the same as the fluctuation exponent
$\gamma_{isom}$ which can be determined from linear-response
experiments at ambient pressure \cite{dittepaper}.  The work is
experimentally heavy, data from 6 different techniques are used,
and it is not easily extended to a large number of samples.

In this paper we develop a prediction from isomorph theory which is
directly related to an experimentally accessible property; the
amplitude of the dielectric loss.  The prediction can be tested with
just two types of relatively standard data: PVT-data in order to
determine an equation of state and high-pressure dielectric
data. Moreover it involves no fitting or scaling parameters.  We show
a test of the prediction on one Van der Waals liquid, 5-phenyl-4-ether
(5PPE), but this can be extended to other samples with a much smaller
effort than any earlier predictions coming from the isomorph theory.

However, there is one experimental challenge; the amplitude of the
dielectric signal has to be measured with high precession.  The
analysis of dielectric data is often focused on the shape and
position of the loss peak, whereas the amplitude is rarely discussed
\cite{russia,highpressure2,highpressure3}.  It is difficult to measure 
the amplitude because it depends on the cell dimension, which often has 
a strong (and sometimes uncontrolled) dependence on temperature 
and pressure.  To overcome this problem we have developed a new cell 
which is capable of giving reliable loss-amplitude data.

The method in this paper is straightforward.  From theory: use the 
fluctuation-dissipation theorem and isomorph theory to predict the 
dielectric-relaxation invariant expression for isomorph states.  By 
experiment: find isochronal states of 5PPE and check if the dielectric 
relaxation obeys the prediction.  We here make use of the fact that 
an isochrone is an isomorph - if the liquid in question is Roskilde 
simple.  The paper is structured as follow.  In sec. \ref{sec:prediction} 
we derive the invariant terms from isomorph theory and develop the 
method of applying the invariance to the measured dielectric signal; 
in sec. \ref{sec:setup} we describe the experimental setup and show 
the result from 5PPE measurement in sec. \ref{sec:result}.

\section{Prediction} \label{sec:prediction} 
In this section we start from the analysis of dipole-moment response 
to an external electric field and its fluctuation.  From the structural 
and dynamic invariance of polarization of isomorph states in reduced 
units we develop four equivalent isomorph-invariant terms, one of 
which is used subsequently to analyze the dielectric signal in our 
measurement.

\subsection{Isomorph prediction}
Suppose an isotropic material consists of $N$ rigid dipolar molecules in 
volume $ V $ with polarization $ \mathbf{P} $ per unit volume.  Then 
the total dipole moment $V \mathbf{P} $ is a vector sum
\begin{equation} 
  V \mathbf{P} =\sum_{i=1}^{N} \mathbf{p}_{i}
\end{equation}
where $\mathbf{p}_{i}$ is the dipole moment of the i$^{\mathrm{th}}$ 
molecule.  In the case where there is no external electric field the 
total sum is zero on average.  When an external electric field $\mathbf{E}$ 
is applied, there is a change in the total average dipole moment 
given \cite{clamp,BroadbandDielectricSpectroscopy,tage} by
\begin{equation} \label{ploarrelax}
  V(t) \mathbf{P} (t) =
  \int_{0}^{\infty} V(t') 
  \mathrm{\epsilon _{0}} \chi _{e} (t') \dot{\mathbf{E}}(t-t') d t'
\end{equation} 
where $\mathrm{\epsilon _{0}}$ is the permittivity of free space, $\chi _{e}$ the 
electric susceptibility of the material, and $\dot{\mathbf{E}}$ the time 
derivative of the field.  For a step change of the external field at 
$ t=0 $ from $ 0 $ to $ E_{0} $ we find
\begin{eqnarray} \label{eq:stepchange}
  V \mathbf{P} (t) 
  &=E_{0}\int_{0}^{\infty} 
  V \mathrm{\epsilon _{0}} \chi _{e} (t') \delta(t-t') d t'\\ \nonumber
	&=E_{0} V \mathrm{\epsilon _{0}} \chi _{e} (t)
\end{eqnarray}
where $\delta$ is the delta function. 

According to the fluctuation-dissipation (FD) theorem one has \cite{
theoryOfDielectrics,ViscoelasticBehavior,polymer}
\begin{equation}
  \int_{0}^{t} \mu (t') d t' = \frac{1}{2 \mathrm{k_{B}} T}  
  \langle (V\mathbf{P}(t)-V\mathbf{P}(0))^{2} \rangle
\end{equation}
where $\mathrm{k_{B}}$ is the Boltzmann constant, $T$ the temperature 
of the material and $ \mu $ the so-called 
memory function, related to the response function of Eq. 
(\ref{eq:stepchange}) by $ V \mathrm{\epsilon _{0}} \chi _{e} (t)=\int_{0}^{t} 
\mu (t') d t' $ \cite{polymer}.  Therefore,
\begin{equation}
  V \mathrm{\epsilon _{0}} \chi _{e} (t) = \frac{1}{2 \mathrm{k_{B}} T} 
  \langle (V\mathbf{P}(t)-V\mathbf{P}(0))^{2} \rangle.
\end{equation}
Defining $ \rho \equiv \frac{N}{V} $ \cite{pressureEnergy4}, and using 
reduced unit for the time t, here $\tilde{t} =t/t_{0}=t 
\rho^{1/3} \sqrt{\mathrm{k_{B}} T/m} $ \cite{pressureEnergy4}, where $m$ is 
the average molecular mass of the material, the above equation becomes
\begin{equation} \label{eq:firststep} 
 \chi_{e}(\tilde{t})= \frac{\rho}{ T }
  \frac{1}{2\epsilon_{0}\mathrm{k_{B}} N} 
  \left<(V\mathbf{P}(\tilde{t})-V\mathbf{P}(0))^{2}\right>.
\end{equation}

The central statement of this paper is that the term $ \langle
(V\mathbf{P}(\tilde{t})-V\mathbf{P}(0))^{2} \rangle $ is preserved
along an isomorph curve.  The argument is simple: along an isomorph 
the structure and dynamics are the same (as function of reduced 
time $\tilde{t}$) - only the intermolecular distances are scaled 
but that does not affect the molecular orientation (note that the 
molecule size is not scaled).  From this isomorph invariance, it is 
possible to derive other isomorph invariant terms which are more directly 
related to what we measure in experiments.  

Since $ 1/(2 \mathrm{k_{B}\epsilon _{0}}  N) $ is constant for a specific
system, it follows from Eq. (\ref{eq:firststep}) that $ \chi_{e}
(\tilde{t}) T/\rho$ is also invariant.

The isomorph theory also states that defining $ \Gamma \equiv 
\rho^{\gamma}/T $, $ \Gamma $ is a constant along any isomorph curve 
where $ \gamma $ is the density-scaling factor \cite{pressureEnergy4,
Stronglycorrelatingliquids}.  Thus, from Eq. (\ref{eq:firststep}) we get
\begin{equation} \label{eq:thirdstep}
 \chi_{e}(\tilde{t})= \frac{1}{ \rho^{\gamma -1} }
 \frac{\Gamma}{\mathrm{2\epsilon_{0}k_{B}} N}
 \langle (V\mathbf{P}(\tilde{t})-V\mathbf{P}(0))^{2} \rangle.
\end{equation}
It follows that $ \chi_{e}(\tilde{t}) \rho^{\gamma-1} $ is predicted 
to be isomorph invariant.  By applying Laplace transformation we also 
obtain the invariance of $ \chi_{e} (\tilde{f}) T/\rho$ and $ \chi_{e}
(\tilde{f}) \rho^{\gamma-1} $ in the frequency domain.  It is the 
latter of these expressions which we use to test the prediction in 
sec. \ref{sec:setup}. 

There is another prediction derived from the original invariance which
connects to classical dielectric theory.  In the Kirkwood-Fr\"{o}lich 
Formula \cite{kirkwood1939,theoryOfDielectrics,BroadbandDielectricSpectroscopy} 
\begin{equation} \label{eq:kirkwoodinrhoT}
  \epsilon_s - \epsilon_\infty 
  = \frac{\mu^2}{ \mathrm{3\epsilon_0  k_{B}} } F g \frac{\rho}{T}
\end{equation}
where ``F'' is the local field correction factor and ``g'' is the correlation 
factor, and $ \mu $ is the permanent dipole moment of each molecule, the 
dielectric loss strength $ \epsilon_s - \epsilon_\infty$.  The dielectric 
loss strength can also be obtained by integrating the imaginary part \cite{
theoryOfDielectrics,TheoryofElectricPolarization,BroadbandDielectricSpectroscopy},
\begin{equation} \label{eq:realandimagconvertor}
\epsilon_s - \epsilon_\infty 
  = \frac{2}{\mathrm{\pi}} \int_{0}^{\infty} \epsilon_{r}''(\tilde{f}) \mathrm{d} \ln \tilde{f}.
\end{equation}

Combining Eqs. (\ref{eq:kirkwoodinrhoT}) and (\ref{eq:realandimagconvertor}), 
and using $\epsilon_{r}''(\tilde{f})=\chi_{e}''(\tilde{f})$, we find 
\begin{equation} 
\frac{\mathrm{\pi} \mu^2}{6 \mathrm{\epsilon_{0} k_{B}} }  F g=  
\int_{0}^{\infty} \frac{T}{\rho} \chi_{e}''(\tilde{f}) \mathrm{d} \ln \tilde{f}.
\end{equation}
It follows that the product of the correction and correlation factor, Fg, is
isomorph invariant.  Thus the prediction actually tells us that the ``Fg'' factor 
should be constant for states along an isomorph curve.  In this prediction 
we specify the state dependence of ``Fg'' is unique along each isomorph curve.

In summary the four isomorph-invariant terms are \[  \left\{ 
  \begin{array}{ll}
  &\langle (V\mathbf{P}(\tilde{t})-V\mathbf{P}(0))^{2} \rangle ; \\
  &\chi_{e} (\tilde{t}) T/ \rho \hspace{1.0cm} 
  ( \mbox{or} \hspace{0.2cm} \chi_{e} (\tilde{f}) T/\rho ) ;\\
  &\chi_{e}(\tilde{t}) \rho^{\gamma-1} \hspace{0.5cm} 
  ( \mbox{or}  \hspace{0.2cm} \chi_{e}(\tilde{f}) \rho^{\gamma-1} );  \\ 
  & \mbox{and} \hspace{0.2cm} Fg \hspace{0.5cm} 
  \mbox{in the Kirkwood-Fr\"{o}lich Formula} .
  \end{array}  
\right. \] 
The first term is used to derive the following three
equivalent isomorph-invariant ones; the second and third can be
further used in loss-peak-amplitude scaling; the last one is added to
make a connection to the classical dielectric theory.  

The result is the dielectric special case of the well
established isomorph prediction that the relaxation function and
relaxation time (in reduced units) are invariant along an isomorph
\cite{pressureEnergy4}. This leads to the prediction of isochronal
superposition, as we discuss in detail in Ref. \cite{lisapaper}.  
The new development is that we now predict exactly how the amplitude
varies along the isomorph.
In the next section we derive the method to 
apply the invariance of $ \chi_{e}(\tilde{f})\rho^{\gamma-1} $ to the data from 
our measurement.

\subsection{From Prediction To Data}
Here we show how the invariance of $ \chi_{e}(\tilde{f})\rho^{\gamma-1} $ can
be tested by experiment.  A similar use  of $ \chi_{e}(\tilde{f}) T/ \rho $ is
presented in the appendix.  For two isochronal
states $(P_1,T_1)$, $(P_2,T_2)$, with density $\rho_1, \rho_2$,
electric susceptibility $\chi_e^{(1)},\chi_e^{(2)}$, and relative
permittivity $\epsilon_{r}^{(1)}(\tilde{f}),\epsilon_{r}^{(2)}(\tilde{f})$, there is the
following relationship $\chi_{e}^{(1)}(\tilde{f})\rho_1^{\gamma-1}=\chi_{e}^{(2)}
(\tilde{f})\rho_2^{\gamma-1}$.  

The measured quantity is the capacitance $C(\tilde{f})$ of a full cell and the
relative permittivity is found by dividing with the empty capacitance
$\epsilon_{r}(\tilde{f})=C(\tilde{f})/C_{\text{empty}}$.  The real part of the relative 
permittivity is affected (and sometimes dominated) by the atomic 
polarization while isomorph theory only predicts the configurational 
polarization.  By inspecting the imaginary part of $\epsilon_r$ we avoid 
the effect from the atomic polarization and we have $\rm Im \left( 
\epsilon_{r}(\tilde{f})\right) =\rm Im \left( \chi_{e}(\tilde{f})\right)$.

Thus the prediction becomes $\rm Im\left(C_{1}(\tilde{f})/C_{\text{empty}}^{(1)} 
\right)\rho_1^{\gamma-1}=\rm Im \left( C_{2}(\tilde{f})/C_{\text{empty}}^{(2)}\right)
\rho_2^{\gamma-1} $.  Because the ambient pressure is more stable than 
elevated pressure, we use the plot of $-C_{1}''(\tilde{f})$ from the 
ambient-pressure measurement to predict the curve $-C_{2}''(\tilde{f})$ 
at elevated pressure, 
\begin{equation} \label{Pred1}
 -C_{2}''(\tilde{f})= -C_{1}''(\tilde{f}) \left(\frac{\rho_1}{\rho_2} \right)^{\gamma-1}
 \frac{C_{\text{empty}}^{(2)}}{C_{\text{empty}}^{(1)}} 
 \hspace{0.5cm} (\text{prediction} ) .
\end{equation}

Eq. (\ref{Pred1}) is later used to treat data from our measurement, and a 
similar equation can be derived from the invariance of $\chi_{e}(\tilde{f}) T/ \rho $ 
(see Appendix).  It predicts the $ -C_{2}''(\tilde{f})$ plot 
at high-pressure state isochronal to the ambient state, if the density data, 
the density-scaling factor of the liquid, and the change of the empty 
capacitance are known.

For isomorph theory the consideration of reduced unit is inevitable in theoretical 
analysis and computer simulations when large changes of density is
possible. However, in physical experiments, because of the relatively small changes of density, 
the influence of reduced unit can be neglected (To see the influence of reduced unit 
on actual data, see Figure 3 in Appendix.).  Therefore, in practice Eq. (11) contains 
isochronal superposition in the classical sense first presented in
Refs. \cite{superposition1,superposition3}, and it moreover predicts 
the scaling between isochronal states beyond normalization for isochronal superposition.

\section{Experimental Setup} \label{sec:setup}
\begin{figure}
 \includegraphics[scale=0.45]{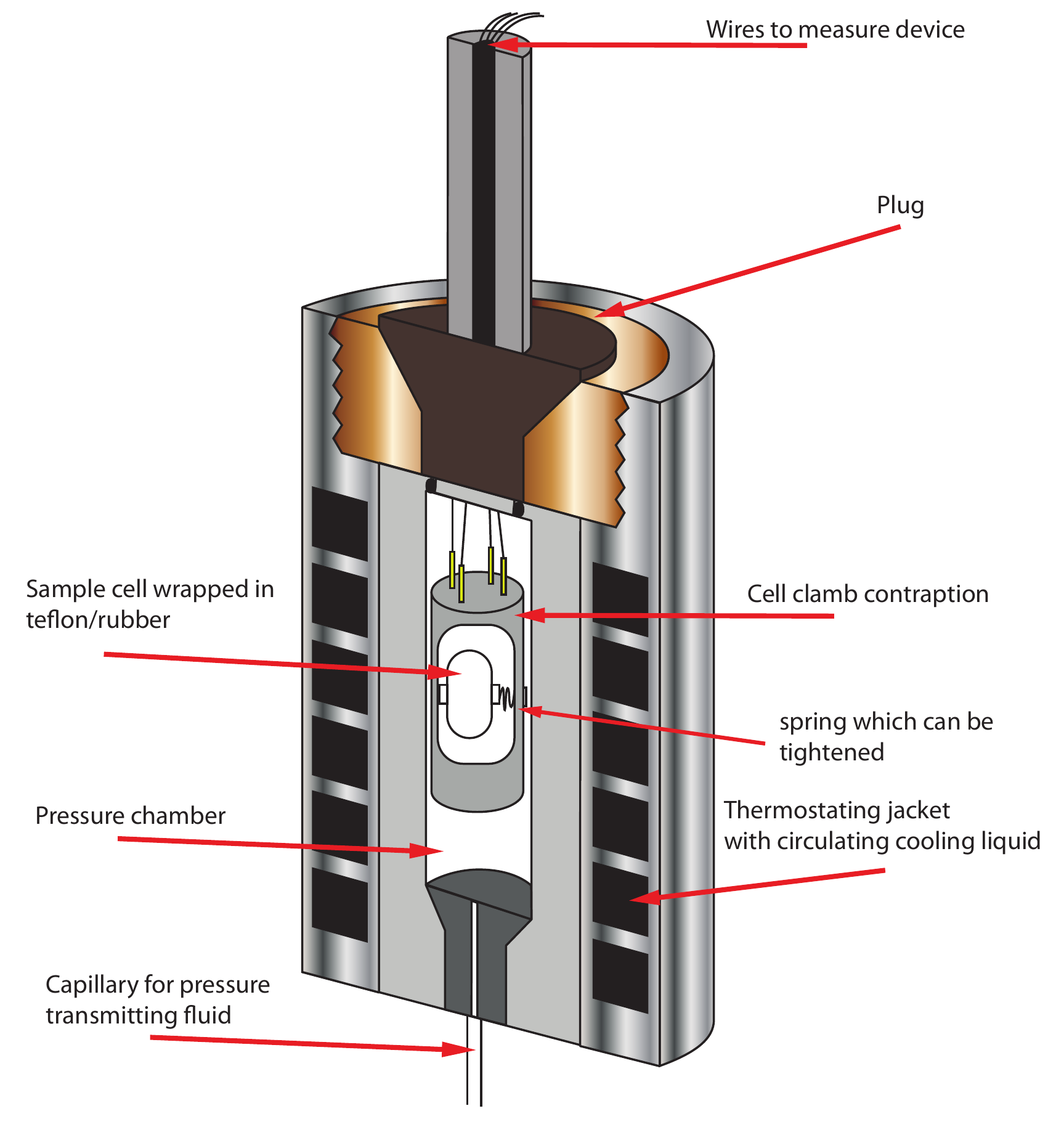}
 \caption{Sketch of the pressure chamber. The chamber has an inner diameter 
 of $30$ mm, inner height of $70$ mm, outer radius of $150$ mm, and the 
 dielectric cell is $20$ mm in diameter.}\label{fig:sketch} 
\end{figure}

The experimental setup consists of a homemade sample cell, a pressure 
chamber connected to a pressure pump (MV1) from Unipress Equipment in 
Warsaw Poland \cite{unipress}, a thermal bath Julabo F81-ME \cite{Julabo} 
and an electrical controlling system that is home made.  The pressure setup 
is described in detail in Refs. \cite{unipress,dittethesis,Lisa} and the 
electrical controlling setup in Refs. \cite{Igarashi2}.  Fig. \ref{fig:sketch} 
is a drawing of the vertical cross section of the pressure chamber.

The thermal liquid, a mixture of 40 $\%$ water and 60 $\%$ ethylene glycol, 
surrounded the pressure chamber, and the thermal couple was about 0.5 cm away 
from the cell.  From $10^{-2} - 10^2$ Hz we measured the capacitance with 
a custom-built voltage generator in combination with an HP 3458A multimeter 
and an HP 4284A LCR meter in the $10^2-10^6$ Hz range. 

The capacitor consisted of two stainless-steel disks each $20$ mm in 
diameter.  To have a large and stable capacitance four square-sapphire spacers 
with dimension $0.5 \text{ mm} \times 0.5 \text{ mm} \times 0.05 
\text{ mm} $ were glued to one of the disks.  After the liquid was 
inserted between the disks (excess liquid was put around the cell gap) 
a Teflon ring and three layers of Teflon tape and latex were wrapped 
around the cell to separate the sample liquid from the pressure liquid.  Since 
the sample liquid might expand and push the two plates further apart upon 
heating \cite{clamp}, a spring force was applied to hold the two plates 
tight.  This was a crucial part in keeping the measured capacitance 
reproducible.  It was then put into a pressure chamber filled with pressure 
liquid (silicon oil) where we were able to control the pressure within 
$\pm 3$ MPa accuracy.  Assuming the glue added no thickness to the spacers, 
this gave an empty capacitance of $55.6$ pF.  However, our measurements 
yielded $51$ pF, suggesting that the size of the glue was not negligible. 

An earlier experiment \cite{Lisa} identified states along two isochronal 
curves, 7 states in total at $0.1$ MPa, $100$ MPa, $200$ MPa and 
$300$ MPa, with loss-peak frequencies around $10^3$ and $10^4$ Hz 
respectively.  In our measurements we also measured nearby isobaric states 
to obtain more isochronal states.  Using the formula for the characteristic 
discharge flow time for pressure-driven channel flow \cite{Lautrup,clamp} 
and isothermal bulk modulus and viscosity data for 5PPE \cite{TinaPaper2013,
TinaThesis}, we estimated the corresponding flow times to be most 60 s and 
6 s.  This meant that even though the spacing was narrow the liquid flowed 
freely and could come into hydrostatic equilibrium on a timescale of a few 
minutes.  To make sure the cell was in thermal equilibrium, we waited at least 
1 hour before each measurement.  We made two measurements at each measuring 
point 15 minutes apart in order to see if there was any change.  The 
reproducibility of data was also checked by repeating the measuring 
protocol, i.e., we increased the pressure to $300$ MPa in steps and dropped 
it symmetrically and then increased again, see details in Ref. \cite{xthesis}, 
where the repeated measurements in the latter two processes showed 
$\pm 0.1 \%$ variation, and these data are used in further analysis. 

It is not trivial to determine the capacitance variation of the empty 
capacitor upon isochronal-state change, and details are discussed in 
Ref. \cite{xthesis}.  Since the specification of the glue upon pressure 
jump is not known and it may undergo plastic deformation above 
100 MPa, we first ignore the glue and only consider the deformation of the 
stainless-steel disks and the sapphires upon temperature and pressure change, 
and the estimated change of the empty capacitance is at most $0.1 \%$ 
within the measuring range.  We further make computer-fitting program 
to estimate the expansion coefficient and the height of the glue, 
and the estimated change of the empty capacitance including the glue 
is about $0.1 - 0.3 \%$ \cite{xthesis}.  Therefore, in the further data 
analysis one approximation is made: for any two isochronal states we 
assume that $C_{\text{empty}}^{(1)}=C_{\text{empty}}^{(2)}$, since the 
predicted amplitude changes we are looking for are of the order $10-20\%$, 
while the estimated change of the empty capacitance is about 
$0.1 - 0.3 \%$ and can be neglected.

\section{Results}\label{sec:result}
\begin{figure}
 \includegraphics[scale=0.45]{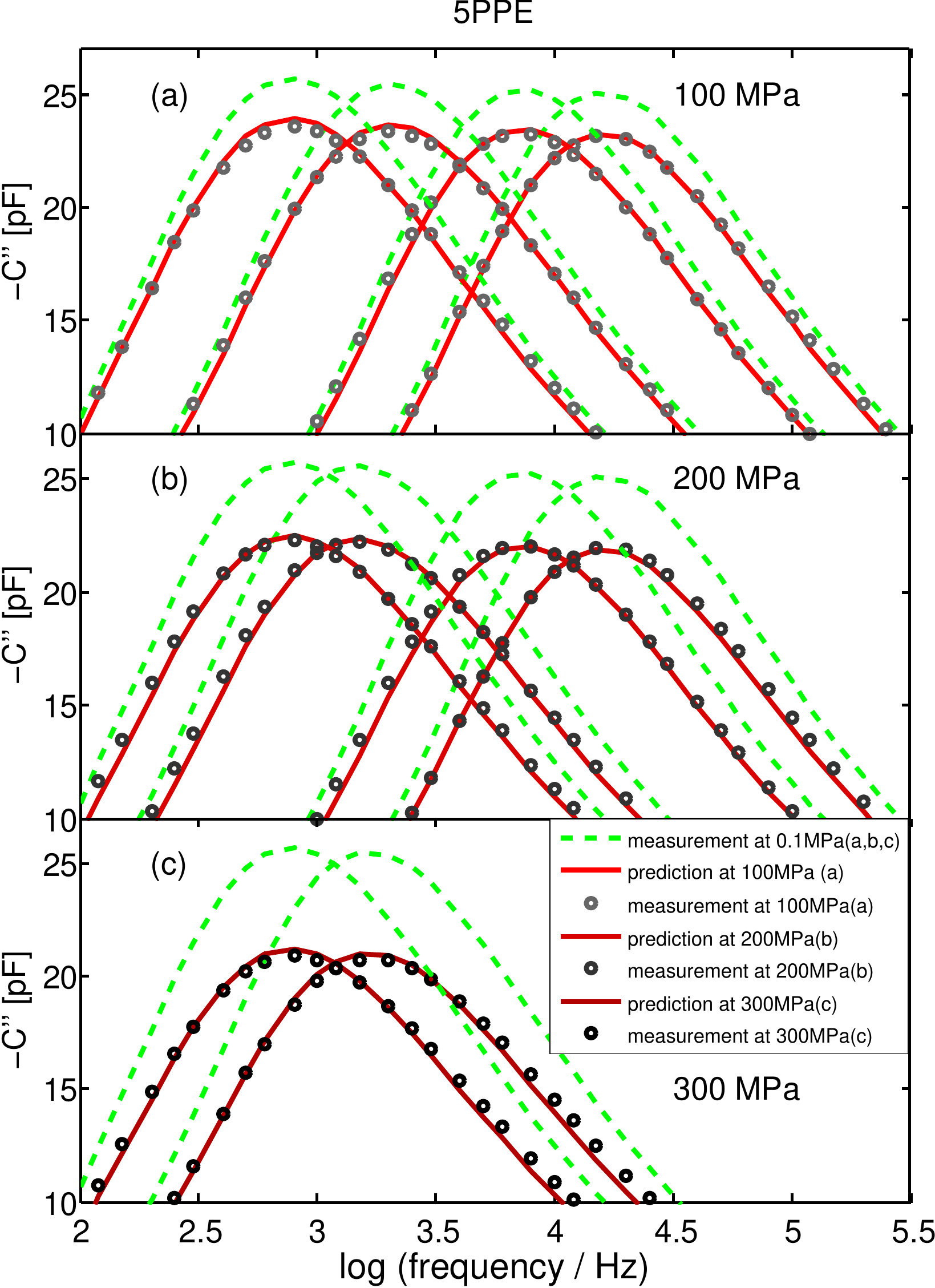}
 \caption{Measurements of $ -C''(f) $ at ambient pressure used to
   predict $ -C''(f) $ of the corresponding isochronal states at $100$
   MPa, $200$ MPa, and $300$ MPa.  The prediction is calculated from
   the isomorph invariance of $ \chi_{e}(f)\rho^{\gamma-1} $ for
   $\gamma=5.5$ \cite{dittepaper}, and compared to the actual
   measurements at elevated pressures. The green dashed lines are
   measurements at ambient pressure, the red curves are the predicted
   high-pressure $ -C''(f) $ plots. There are no fitting parameters or
   arbitrary scaling involved in this prediction. The black circles
   are the high-pressure measurements that should coincide with the
   red curves. The 10 pairs of isochronal states in this figure are
   selected from 42 pairs in total such that they can be distinguished
   as different isochronal states visually.  The empty cell
   capacitance is $51$ pF at ambient pressure.  Notice that the plots
   shown here are on a linear scale. The log-log plot scaled with the
   peak position are shown in Fig. \ref{fig:reducedANDnotreduced} in
   the Appendix along with a short discussion on isochronal
   superposition.  }
 \label{fig:mainfigure} 
\end{figure}

Fig. \ref{fig:mainfigure} is a visual presentation of the prediction 
( Eq. (\ref{Pred1}) ).  The density of 5PPE is found by extrapolation 
of the Tait equation whose parameters have been fixed by another 
experiment \cite{dittethesis}.  The density-scaling factor $\gamma=5.5$ 
is determined by a previous experiment \cite{dittepaper,dittethesis}.  In 
total we have measured 155 state points, and found 42 pairs of 
isochronal states.  The figure shows 10 pairs of isochronal states that 
have loss-peak frequencies spread evenly.  Because we assume that 
$C_{\text{empty}}^{(1)}=C_{\text{empty}}^{(2)}$ and the prediction is 
formulated in terms of capacitance, Fig. (\ref{fig:mainfigure}) plots 
$ -C''(f) $ instead of the dielectric constant $\epsilon_{r}$.  Within 
our measuring range the dielectric loss strength $ \epsilon_{s}-
\epsilon_{\infty}$ obtained is within $ 1.2-1.6 $.  
In Fig. \ref{fig:mainfigure} we see overall good match between the predicted plots of 
$ -C''(f)$ (red curves) and the actual measurements (black circles) at 
all three elevated pressures. This is impressive because there are no
fitting parameters or arbitrary scaling parameters involved. 
 The data here is limited to one sample 
and also covers a limited range of pressure and temperature.  Nevertheless 
it provides direct experimental support to the isomorph theory. 

\section{Summary}
The isomorph prediction for dielectric relaxation along isomorph curve
has been derived by means of the theory of dielectrics, linear-response 
theory and the FD theorem.  Based on the isomorph invariance of the 
structural and dynamic configuration of Roskilde simple liquid, four 
equivalent invariant terms in dielectric relaxation are obtained.  We 
have moreover developed the details needed to use one of these terms 
to analyze data.

The homemade capacitor cell has produced reliable data.  By filling
the cell with liquid 5PPE we have measured at isochronal states, and the
measurement at ambient pressure is used to predict that of elevated
pressure.  Comparing the negative-imaginary-capacitance spectroscopy
from the prediction and measurement at elevated pressures, we have seen
agreement between the two plots, providing support to the isomorph
theory.

The ambition is that the prediction presented in this paper can be
used to test the isomorph theory on several liquids by us and even
more importantly by other groups doing high-pressure dielectrics.  

\begin{acknowledgments}
We thank Lisa A. Roed for helping with isochronal-superposition treatment 
and comparison with her data.  
The centre for viscous liquid dynamics ``Glass and Time'' is sponsored 
by the Danish National Research Foundation via grant DNRF61. Kristine Niss 
wishes to acknowledge The Danish Council for Independent Research for supporting this work.
\end{acknowledgments}

\appendix
\setcounter{secnumdepth}{0}
\section{Appendix}\label{sec:Appendix}
Similar to Eq. (\ref{Pred1}), we derive the following equation based 
on the invariant term $ \chi_{e}(f) T/ \rho $, 
\begin{equation} \label{Pred2}
 -C_{2}''(f)= -C_{1}''(f)\frac{ T_1 \rho_2}{T_2 \rho_1} 
 \frac{C_{\text{empty}}^{(2)}}{C_{\text{empty}}^{(1)}}. 
\end{equation}
The two equations are considered equivalent from isomorph theory.  
\begin{figure}[h!]
 \includegraphics[scale=0.65]{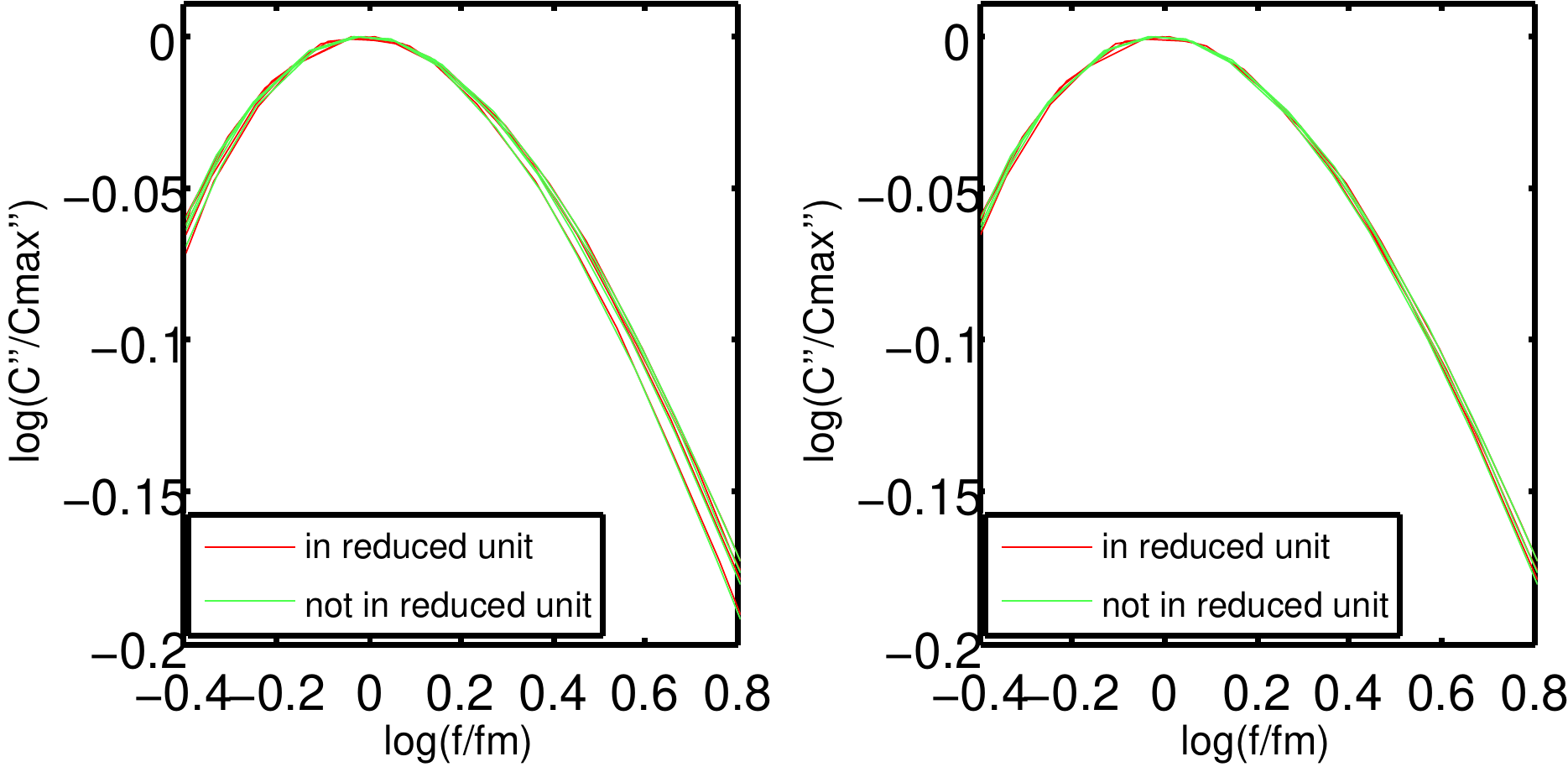}

 \caption{A conventional isochronal-superposition examination of our 5PPE data 
 at loss-peak frequency around $10^{3}$ Hz, plotting the normalized imaginary capacitance 
 by its amplitude $C''_{\mathrm{max}}$ against normalized frequency by the loss-peak frequency $f_{\mathrm{m}}$.  
 The red curves are made for isochronal states in reduced units and the green ones not 
 in reduced units.  The plots on the left are plotted for states at 0.1 MPa, 100 MPa, 200 MPa 
 and 300 MPa, while the plots on the right are at 100 MPa, 200 MPa and 300 MPa, to show that 
 isochronal superposition is better obeyed for states at elevated pressures.
 }\label{fig:reducedANDnotreduced} 
\end{figure}

\begin{figure}
 \includegraphics[scale=0.5]{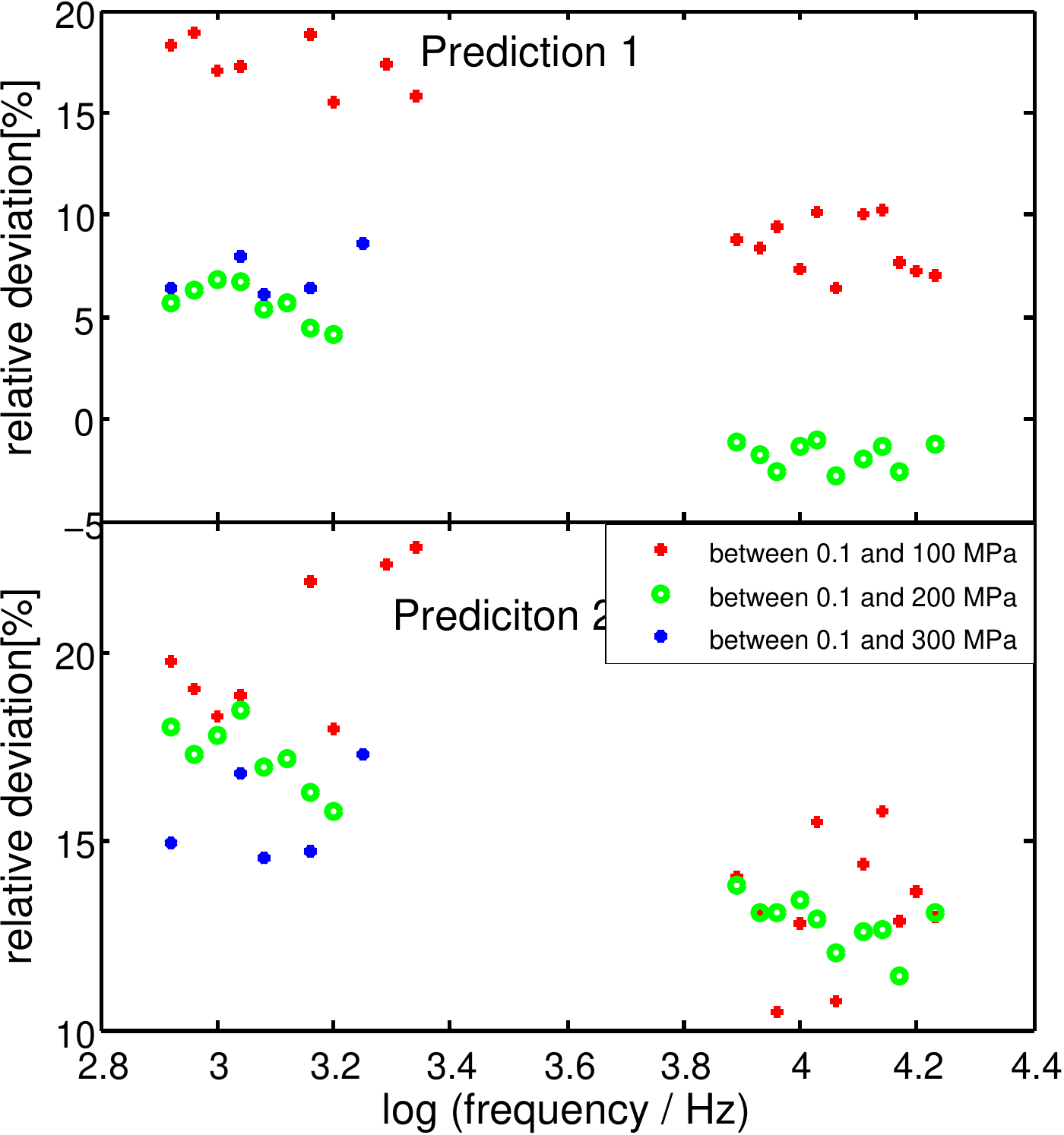}
 \caption{Calculated ``relative deviation'' for both predictions.  The 
 ``relative deviation'' is defined as the ratio of differences in 
 $ -C''(f) $ amplitudes between the difference of the predicted curve 
 and the actual measurement and the difference of the ambient pressure 
 and high-pressure measurements.  (a) is made for isomorph invariance of 
 $ \chi_{e}(f)\rho^{\gamma-1} $ - Prediction 1 (Eq. (\ref{Pred1})) and 
 (b) for isomorph invariance of $ \chi_{e}(f) T/\rho $ - Prediction 2 
 (Eq. (\ref{Pred2})).  For Prediction 1 the relative deviation is within 
 $(-3 , 19 )\%$, and for Prediction 2 it is within $(10 , 24 )\%$.}
 \label{fig:reP}
\end{figure}

Now that a visual check (Fig. \ref{fig:mainfigure}) supports the
isomorph theory, we further show a way to quantify the degree of
agreement between the prediction and measurement at elevated
pressures.  We have developed a measure called the ``relative
deviation''.  

It has been shown 5PPE obeys isochronal superposition to a high degree \cite{lisapaper,Lisa}.  
We have also done similar isochronal-superposition examination to our 5PPE 
data.  The plot from Fig. \ref{fig:reducedANDnotreduced} tells that 5PPE 
obeys isochronal superposition well, especially for states at elevated pressures.  
In addition there is no distinction in analyzing isochronal superposition 
of isochronal states in reduced unit or not. 
Thus it suffices to exploit the loss-peak amplitude of a $-C''(f)$ plot.

Our measure 
takes into account of the loss-peak amplitude $A_0$ of $-C_{1}''(f)$, 
the amplitude $A_p$ of the predicted $-C_{2}''(f)$ plot, and the 
measured amplitude $A_m$ of the measured $-C_{2}''(f)$ plot, and 
calculate the ratio $\delta = ( A_p-A_m) / ( A_0-A_m)$.  The smaller 
$\delta$ is the better the isomorph prediction works.  From the 42 
pairs of isochronal states we find that the values of relative 
deviation is within $  -3 - 19 \%$ when applying the $ 
- \epsilon_{r}''(f) \rho^{\gamma-1} $ prediction and within 
$ 10 -24 \%$ for the second prediction.  Figure (\ref{fig:reP}) 
plots the relative deviation against loss-peak frequencies in 
a logarithmic scale for both predictions. 

\end{document}